\documentclass{birkjour}
 \theoremstyle{definition}
 
 \theoremstyle{remark}

 \numberwithin{equation}{section}

\begin{document}

%
%
%
%
%
%
%
%
%

\title{Iterants, Idempotents and Clifford algebra in Quantum Theory}

\author{Rukhsan Ul Haq}

\address{%
Theoretical Sciences Unit,\\
Jawaharlal Nehru Center for Advanced Scientific Researcch\\
Jakkur Bangalore India}

\email{mrhaq@jncasr.ac.in}

\author{Louis H. Kauffman}

\address{
Department of Mathematics and Statistics\\
University of Illinios\\
Chicago USA.}

\email{kauffman@uic.edu}


\date{}

\begin{abstract}
Projection operators are central to the algebraic formulation of quantum theory because both wavefunction and hermitian operators(observables) have spectral decomposition in terms of the spectral projections. Projection operators are hermitian operators  which are idempotents also. We call them quantum idempotents. They are also important for the conceptual understanding of quantum theory because projection operators also represent observation process on quantum system. In this paper we explore the algebra of quantum idempotents and show that they generate Iterant algebra (defined in the paper), Lie algebra, Grassmann algebra and Clifford algebra  which is very interesting because  these later algebras were introduced for the geometry of spaces and hence are called geometric algebras. Thus the projection operator representation gives a new meaning to these geometric algebras in that they are also underlying algebras of quantum processes and also they bring geometry closer to the quantum theory. It should be noted that projection operators not only make lattices of quantum logic but they also span projective geometry. We will give iterant representations of framed braid group algebras, parafermion algebras and the $su(3)$ algebra of quarks. These representations are very striking because iterant algebra encodes the spatial and temporal aspects of recursive processes. In that regard our representation of these algebras for physics opens up entirely new perspectives of looking at fermions,spins and parafermions(anyons).

\end{abstract}

\maketitle

\section{Introduction}
Dirac in his monumental work\cite{Dirac} introduced his famous notation known as {\it bra-ket} notation in quantum theory. This notation has become integral part of quantum theory formalism. Dirac bras and kets are algebraic objects and are basis independent. Bras and kets are very abstract elements of formalism of quantum theory. One can choose a basis and find representations of kets and bras in the given basis. In  state space of the given quantum system  kets are the vectors and bras belong to the dual space. Dirac bra-ket notation hides underneath a rich underlying algebra of quantum theory. So Dirac bra-ket notation is not just notation, rather it has in it the  algebra of quantum theory. When taken  abstractly, Dirac bra-ket quantum algebra also  goes beyond the domains where it arose into the biology of reproduction of DNA \cite{Kauffman1}. Its generalization in the form of Temperley-Lieb algebra has become a central algebraic structure for integrability of statistical models \cite{Baxter},Topological Quantum Field Theory (TQFT), knot invariants \cite{Kauffman2} and  quantum groups\cite{Kauffman3} which form a very fertile ground of mathematics and physics and are related to all modern theories of physics and mathematics. So right at the beginning of quantum theory in Dirac's beautiful  bra-ket formalism lies an elegant quantum algebra which finds realizations in all fields of modern physics and mathematics. We are going to explore some aspects of this quantum algebra that are related to fermions, and more generally to Clifford algebras and Iterant algebras. Iterant algebra which arises from recursive structures in mathematics and from the role of permutations in matrix algebra. Bringing these algebras together sheds new light on conceptual paradoxes that quantum theory poses: the questions of observer and time being two very important ones. \\

There are two special algebraic elements that are central to quantum physics. One is the quantum amplitude for going from one quantum state to another quantum state.  The amplitude  is given by $\langle\psi\mid O \phi\rangle$ where $O$ is any observable of quantum system. Here we use inner product of Dirac bra-ket algebra. Using the outer product in this algebra we can form another very important element of the algebra which is an idempotent written as $\mid\psi\rangle \langle \psi\mid$. We will call it this a  "quantum idempotent". Wheras the quantum amplitude squared gives the probability for a quantum processes to occur, quantum idempotents have all the spectral information of quantum observables and also represent measurements on quantum systems, and hence are central to the formalism of quantum theory. The algebra of quantum idempotents was studied in detail by Schwinger \cite{Schwinger} when he was looking at  the theory of microscopic measurements on quantum systems. Based on this algebra he developed a full formalism for finite quantum systems. This formalism
finds extensive applications in quantum computing(see \cite{Vourdas} and references therein). Yet another direction in which quantum idempotents become very important is the taken by Bohm and Hiley where they find that quantum idempotents represent the process algebra of quantum mechanics \cite{Hiley}. It needs to be mentioned that it is actually Herman Weyl who realized the importance of idempotents in quantum theory \cite{Weyl}. 

In this paper we will study quantum idempotents and explore the algebras they generate. First we will show that quantum idempotents generate  the Grassmann algebra of fermions and Clifford algebra as well. We will in particular study the Clifford algebra of Spin and its $su(2)$ algebra. These algebraic results give new insights about these algebra and open up new perspectives in understanding fermions and their quantum physics.It also gives an answer to question of Arthrur Eddington \cite{Clive} who asked whether fermions are idempotents rather than nilpotents. It is a foundational shift in the understanding of fermions if they are taken to be idempotents rather than nilpotents as is done usually. Quantum idempotents also generate Lie algebras which encode the symmetries of quantum dynamics. These symmetries are different from the geometrical symmetries because nowhere do we use the invariance of the metric of the state space of quantum system, and hence these symmetries are entirely symmetries of quantum dynamics and hence are called dynamical symmetries \cite{Barut}.  Idempotents generate dynamical symmetries of a quantum system.\\

   The rest of this paper is organized in following sections: In section 2 we introduce quantum idempotents and their algebra.  We then look into the small but very fascinating world of a pair of quantum idempotents. This small quantum world constitutes what are called qubit systems. These are the quantum systems with a two dimensional Hilbert space.  We also generalize projection operators and introduce two new  elements of the algebra. The quantum idempotents of qubit systems lets us to explore fermions and spin and we come up with new representations of well-known algebra of these systems. We particularly study Majorana fermions which are the promising candidates for topological quantum computing and see how their Clifford algebra can be generated by quantum idempotents.\\

 Here is a very brief introduction to the concept of an {\it iterant}. Sections 5 to 11 of this paper are devoted to using this concept as we will explain below. An iterant is a sum of elements of the form $$[a_1,a_2,...,a_n] \sigma$$ where $[a_1,a_2,...,a_n]$ is a vector of elements that are scalars and $\sigma$ is a permutation on $n$ letters.
Such elements are themselves sums of elements of the form  $$[0,0,...0,1,0,...,0] \sigma = e_{i} \sigma$$ where the $1$ is in the $i$-th place. The elements $e_{i}$ are the basic idempotents that 
generate the iterants with the help of the permutations. If $a = [a_1,a_2,...,a_n]$, then we let $a^{\sigma}$ denote the vector with its elements permuted by the action of $\sigma.$ 
If $a$ and $b$ are vectors then $ab$ denotes the vector where $(ab)_{i} = a_{i} b_{i}.$
Including permuations, we define iterant multiplcation by the rule $$(a\sigma) (b \tau) = (a b^{\sigma}) \sigma \tau$$ where vectors are multiplied as above and we take the usual product of the permutations.
All of matrix algebra and more is naturally represented in the iterant framework. For example, if $\sigma$ is the order two permutation of two elements, then $[a,b]^{\sigma} = [b,a].$ We can define
$$i = [1,-1]\sigma$$ and then $$ i^2 = [1,-1]\sigma [1,-1] \sigma = [1,-1] [1,-1]^{\sigma} \sigma^{2} = [1,-1][-1,1] = [-1,-1] = -1.$$ In this way the complex numbers arise naturally from iterants.
One can interpret $[1,-1]$ as an oscillation between $+1$ and $-1$ and $\sigma$ as denoting a temporal shift operator. The $i = [1,-1]\sigma$ is a time sensitive element and its self-interaction has square minus one.
In this way iterants can be interpreted as a formalization of elementary discrete processes. \\

 In section 5 we introduce the algebra of iterants and show how they are related to idempotents.  In sections 6 and 7 we show how iterants algebras can represent matrix algebras in a variety of ways. In particular
 we show how iterants based on the group of order two correspond to $2 \times 2$ matrices and how, given a finite group, the multiplication table of the group gives a full matrix algebra generated by the Cayley's permutation
 representation of the group. The quaternions are an interant algebra based on the Klein Four Group. The complex numbers and elementary Clifford algebras arise immediately for the cylic group of order two.
 In section 8 we discuss the framed (Artin) braid group and show how it has a significant iterant representation. We use this representation to construct an iterant algebra corresponding to Sundance Bilson-Thomposn's
 framed braid model for Fermions \cite{Sundance}. We present the iterant representation of parafermions in section 9. In section 10 we give an iterant model for the $su(3)$ Lie algebra and in section 11 we use it to show that Sundance Bilson-Thompson's model can be {\it embedded} in $su(3)!$  This relationship of the framed braids model for Fermions and the $su(3)$ Lie algebra is the end of this paper and it will form the beginning
 of our sequels to the present paper. Section 12 is a discussion about further directions for this research.\\
 

\section{Quantum Idempotents}
Dirac introduced the notation for state vectors and their duals which are kets and bras respectively. Kets and bras can be thought to belong to a state space but here we will take them as abstract algebraic objects. These kets and bras have two types of products: 1. $\langle\psi\mid\phi\rangle$ and 2. $\mid\psi\rangle\langle\phi\mid$ the first one is a scalar and gives expectation value or probability amplitude and the second one is the projection operator.This projector operator satisfies:
\begin{align}
&P=\mid \psi \rangle \langle \psi \mid
& P^{2}=P
\end{align}
For a n dimensional basis we will have n projection operators corresponding to each eigenvalue(we do not consider spectral degeneracies here).
We can write both state vector as well as operators in terms of projection operators.
\begin{equation}
\mid \Psi \rangle = \sum_{i}^{n} \mid P_{i}\Psi\rangle
\end{equation}
Similarly an observable can be written in terms of projection operators.
\begin{equation}
H=\sum_{i}e_{i}P_{i}
\end{equation}
where $e_{i}$ is the i-th eigenvalue of $H$ and $P_{i}$ projects onto the i-th eigensubspace of H.\\
Idempotents encode two very important properties of the eigenfunctions of the hermitian operator very elegantly.
\begin{align}
& \sum_{i} P_{i} = 1  & \text{(Completeness)}\\
& P_{i}^{2}=P_{i}     &  \text{(Idempotence)}\\    
& P_{i}P_{j}=\delta_{ij}P_{j} & \text{(Orthogonality)}
\end{align}
Quantum idempotents become important in quantum theory because for their role in the spectral decompositions of hermitian operators(observables). They also provide the formalism of measurement theory as first developed by Schwinger. There are also approaches to quantum mechanics like as Bohm and Hiley's where quantum idempotents provide the algebra of processes. Though the algebra of quantum idempotents is very simple we will show in this paper that quantum idempotents generate not only Lie algebras but Grassmann and Clifford algebra as well. Also as mentioned in introduction that quantum idempotents also generate Temperley-Lieb algebra and have provided the formalism to understand DNA replication in a biological systems which have seemingly no relation to quantum mechanics where the quantum idempotents were introduced. We believe that understanding quantum idempotents better will give us clues to the resolve the conceptual paradoxes of  quantum theory.


\section{Tale of two projections:Pauli matrices and Fermions}
Lets divide the state space in two subspaces using P and Q projection operators.One space we will call as "Model space" or P-Space and its complement we will call as "Target space" or Q-Space.
\begin{align}
&P^{2}=P \quad Q^{2}=Q \quad PQ=QP=0 \quad P+Q=1
\end{align}
Now lets us consider a quantum system with two states only. Such two level systems are very common in quantum physics and also called  qubit systems in quantum computing. Spin-$\frac{1}{2}$ system being the well known example. We denote its states as $\mid \uparrow\rangle$ and $\mid \downarrow \rangle$. The correspoding projectors P and Q are: $P=\mid \uparrow \rangle \langle \uparrow\mid$ and $ Q= \mid \downarrow \rangle \langle \downarrow \mid$. We can also define two more operators R and S, which are generalized projection operators which satisfy $Z_{2}$ graded superalgebra. Let $U^{pq}=\mid p\rangle \langle q\mid$ and $U^{rs}=\mid r\rangle \langle s\mid$ be two generalized projection operators,they close to the following algebra under multiplication:
\begin{align}
U^{pq}U^{rs}=\delta_{qr}U^{ps}
\end{align}
For the generalized projection operators R and S this algebra leads to following relations:
\begin{align}
&R=\mid \uparrow\rangle \langle\downarrow \mid \quad S=\mid \downarrow \rangle \langle \uparrow\mid \\
& R^{2}=S^{2}=0 \quad R^{\dag}=S \quad [R,S]=P-Q
\end{align}


Now we will show that these projection operators actually can generate $su(2)$ algebra which is very surprising because in non-relativistic quantum mechanics, spin is put in an adhoc manner and it is only in Dirac theory of electron that spin appears naturally in the spinor solution to Dirac equation\cite{Messiah}.
We will write the idempotents explicitly:
\begin{align}
P=\mid \uparrow \rangle \langle \uparrow\mid \qquad Q= \mid \downarrow \rangle \langle \downarrow \mid
\end{align}
\begin{align}
R=\mid \uparrow \rangle \langle\downarrow\mid \qquad S= \mid \downarrow \rangle \langle \uparrow \mid 
\end{align}
The generators of $su(2)$  algebra are 
\begin{align}
&\sigma_{z}=P-Q \qquad \sigma_{x}=\frac{1}{2}(R+S) \qquad \sigma_{y}=\frac{1}{2i}(R-S)
\end{align}
where $\sigma_{x}$, $\sigma_{y}$ and $\sigma_{z}$ are Pauli matrices.
Now using the standard relation between Pauli matrices and quaternions we see that there is quaternionic structure underlying these projection operators of two qubit systems.

\subsection{Representation of Fermion algebra}
In this section we will give a new representation of fermion algebra. The standard fermion algebra is introduced in quantum mechanics at the second quantization level and assumes a special state called vacuum state\cite{JJ}. Here we will see that fermion algebra is already sitting inside the algebra of quantum idempotents.

Let us consider a single fermion system. It has two states which we denote by $\mid0\rangle$
and $\mid1\rangle$. The projection operators corresponding to these two states are
\begin{equation}
P_{0}=\mid0\rangle\langle0\mid \qquad P_{1}=\mid1\rangle\langle1\mid 
\end{equation}
These two projection operators satisfy the algebra\\
\begin{align}
P_{0}^{2}=P_{0} \quad P_{1}^{2}=P_{1} \quad P_{0}P_{1}=P_{1}P_{0}=0\quad
P_{0}+P_{1}=1
\end{align}
Two more generalized projection operators can be introduced:
\begin{align}
P_{01}=\mid 0><1\mid \quad P_{10}=\mid 1><0\mid
\end{align}
We will denote $P_{01}$ and $P_{10}$ by R and S respectively.
They satisfy the algebra
\begin{align}
R^{2}=S^{2}=0 \quad RS=P_{0} \quad SR=P_{1}
\end{align}
The operators R and S satisfy fermion algebra and hence give a representation of fermion algebra.
We can make a mapping from this algebra to fermion algebra.
\begin{align}
S\equiv c \quad R\equiv c^{\dag}
\end{align}
where $c$ and $c^{\dag}$ are fermion annihilation and creation operators respectively.
 R and S satisfy all properties of the fermion operators.
\begin{align}
\{S,R\}=SR+RS= P_{1}+P_{0}=1 \quad \{S,R\}=\{c,c^{\dag}\}=1 \\
S\mid0\rangle=\mid0\rangle\langle1\mid0\rangle=0 \quad S\mid1\rangle=\mid 0\rangle\langle1\mid1\rangle=\mid0\rangle\\
R\mid0\rangle=\mid1\rangle\langle0\mid0\rangle=\mid 1\rangle \quad R\mid1\rangle=\mid 1\rangle\langle1\mid0\rangle=0
\end{align}
We can also define number operator as $N=c^{\dag}c=RS$
\begin{align}
RS\mid1\rangle = \mid1\rangle \quad RS \mid 0\rangle=0 \quad (RS)^{2}=RS
\end{align}

Hence we have presented a new representation of the fermion algebra in generalized projection operators. The relation to the standard Fock space formulation of Fermions is clear.In our representation we have not invoked second quantization but still have algebra of fermions. It shows that algebra of fermions is not tied to second quantization and fermions can be understood in other ways as well. We empahsize that in our representation \textit{fermions are more like processes than as objects}. Yet another important feauture of our representation of fermions as idempotents is that fermion algebra is sub-algebra of algebra of quantum idempotents which has a bosonic subalgebra as well.Superalgebra of quantum idempotents can help us to get insight into the supersymmetry as well.\\
 We stop here to understand this new representation. It offers new perspective about the nature fermions. Algebraically fermions  are generalized projection operators.
So fermions can be thought as quantum processes which take the fermionic system from one state to another. This picture of fermions brings out new aspects of fermions which are not obvious in the Fock space approach to fermions in which fermions are associated with two non-hermitian creation and annihilation operators acting on the vacuum state. Our representation is closer to Eddington's approach\cite{Clive} in which he had got this insight into fermions that they as "elements of existence" should be idempotents rather than nilpotents. Idempotents have the property that they stay stable even after getting transformed. Idempotents are invariants of dynamics. They remain invariant under the dynamical transformantions. So our approach and this new representation opens up new perspectives in understanding the dynamics of fermions.

It is very interesting to see the quantum idempotents also satisfy Grassmann and Clifford algbera which were first  introduced as algberas of geometry \cite{Lounesto}. Here in this context one can relate these algebras to the underlying geometry of the projection operators which is projective geometry.\\
Another curious aspect of this formalism is that it is also related to  iterant algbera which captures the spatial and temporal aspects of the recursive proccesses both in logic and space-time. In later section we will explore this connection of idempotents and iterants in more detail.
\section{Quantum idempotents and Clifford algebra}
Now we will go ahead to show that  quantum idempotents also generate Clifford algebra.  We can combine these $R$ and $S$ operators of previous section to define Majorana fermions which satisfy Clifford algebra.
\begin{align}
R+S = \eta_{1} \quad i(R-S)= \eta_{2}
\end{align}
\begin{align}
\eta_{1}^{2}=\eta_{2}^{2}=1 \\
\{\eta_{1},\eta_{2}\}=0
\end{align}
which can be written more compactly as:
\begin{align}
&\lbrace\eta_{i},\eta_{j}\rbrace=2\delta_{ij}
\end{align}
 $\eta_{1}$ and $\eta_{2}$ are Majorana fermion operators which are obtained from the Dirac fermions as linear combinations. Majorana fermions are presently an active topic of research in topological quantum computing and hence our representation opens up a new way of looking at Majorana fermions. It is very interesting here to see how Kauffman\cite{Kauffman4} has also arrived at the algebra of Majorana fermions from a side of logic in the form of calculus of distinctions. In his representation Majorana fermion becomes a logical particle which follows laws of form. But what is quite profound is that in those logical rules is hidden the fusion algebra of Majorana fermions as anyons. Since quantum idempotents and  iterants are not independent and in fact iterants can be written in terms of idempotents so these two approaches and representaions are related to each other even though they give totally different perspectives of Majorana fermions. In our case fermions are not tied to any Fock space and in fact in our representation number operator is not very important which is the case in the Fock space. So fermion needs not be to be interpreted as some kind of particle rather it is a quantum process. In our approach all that matters is the algebra of projection operators and from that ground arise  fermions,their dynamics and statistics. 
\section{Iterants and Idempotents}
An iterant is a sum of elements of the form $$[a_1,a_2,...,a_n] \sigma$$ where $[a_1,a_2,...,a_n]$ is a vector of elements that are scalars (usually real or complex numbers) and $\sigma$ is a permutation on $n$ letters.
Such elements are themselves sums of elements of the form  $$[0,0,...0,1,0,...,0] \sigma = e_{i} \sigma$$ where the $1$ is in the $i$-th place. The elements $e_{i}$ are the basic idempotents that 
generate the iterants with the help of the permutations.\\

Note that if $a = [a_1,a_2,...,a_n]$, then we let $a^{\sigma}$ denote the vector with its elements permuted by the action of $\sigma.$ 
If $a$ and $b$ are vectors then $ab$ denotes the vector where $(ab)_{i} = a_{i} b_{i},$ and $a + b$ denotes the vector where $(a+b)_{i} = a_i + b_i .$
Then 
$$(a\sigma) (b \tau) = (a b^{\sigma}) \sigma \tau,$$
$$(ka)\sigma = k(a\sigma)$$ for a scalar $k$, and 
$$(a + b)\sigma = a\sigma + b\sigma$$ where vectors are multiplied as above and we take the usual product of the permutations.
All of matrix algebra and more is naturally represented in the iterant framework, as we shall see in the next sections.\\

For example, if $\eta$ is the order two permutation of two elements, then $[a,b]^{\eta} = [b,a].$ We can define
$$i = [1,-1]\eta$$ and then $$ i^2 = [1,-1]\eta [1,-1] \eta = [1,-1] [1,-1]^{\eta} \eta^{2} = [1,-1][-1,1] = [-1,-1] = -1.$$ In this way the complex numbers arise naturally from iterants.
One can interpret $[1,-1]$ as an oscillation between $+1$ and $-1$ and $\eta$ as denoting a temporal shift operator. The $i = [1,-1]\eta$ is a time sensitive element and its self-interaction has square minus one.
In this way iterants can be interpreted as a formalization of elementary discrete processes.\\

Note that we can write $a=[1,0], b = [0,1]$ and $A = a \eta, B = b \eta$ where $\eta$ denotes the transpsition so that $[x,y]\eta = \eta[y,x]$ and $\eta^2 = 1.$ Then we have
$$aa = a, bb = b, ab = 0, a + b = 1, AA = 0 = BB, AB = a, BA = b.$$ This is the mixed idempotent and permutation algebra for $n=2.$ Then we have
$$i = A - B$$ as we can see by 
$$ii = (A-B)(A-B) = AA -AB-BA+BB = -a-b = -1.$$ This is the beginning of the relationships between idempotents, iterants and Clifford algebras.\\

Note that we construct an elementary Clifford algebra via
$$\alpha = [1,-1] = a - b$$
and
$$\beta = \eta.$$
Then we have$$ \alpha^2 = \beta^2 = 1$$ and $$\alpha \beta + \beta \alpha = 0.$$
Note also that the non-commuting of $\alpha$ and $\beta$ is directly related to the interaction of the idempotents and the permutations.
$$\alpha \beta = [1,-1] \eta = \eta [-1, 1] = - \eta [1,-1] = - \beta \alpha.$$\\

\noindent Iterant algebra is generated by the elements $$e_{i} \sigma$$ where $e_{i}$ is a vector with a $1$ in the $i$-th place and zeros elsewhere, and $\sigma$ is an abritrary element of the symmetric group
$S_{n}.$ We have that $$e_{i} \sigma = \sigma e_{\sigma^{-1}(i)}$$ so that the multiplication of iterants is defined in terms of the action of the symmetric group. We have
$$e_{i} \sigma e_{j} \tau = e_{i} e_{\sigma(j)} \sigma \tau = \delta(i, \sigma(j)) e_{i}  \sigma \tau.$$
By themselves, the elements $e_{i}$ are idempotent and we have $$1 = e_{1} + \cdots e_{n}.$$ The iterant algebra is generated by these combinations of idempotents and permutations.\\

 \section{MATRIX ALGEBRA VIA ITERANTS}
 We translate iterants to matrices as follows. We write
 
 $$[a,b] + [c,d]\eta = 	\left(\begin{array}{cc}
			a&c\\
			d&b
			\end{array}\right).$$ 

\noindent where
$$[x,y] = 	\left(\begin{array}{cc}
			x&0\\
			0&y
			\end{array}\right).$$ 

\noindent and
$$\eta = 	\left(\begin{array}{cc}
			0&1\\
			1&0
			\end{array}\right).$$ 

Recall the definition of matrix multiplication.
$$\left(\begin{array}{cc}
			a&c\\
			d&b
			\end{array}\right)
			\left(\begin{array}{cc}
			e&g\\
			h&f
			\end{array}\right) =
			\left(\begin{array}{cc}
			ae+ch&ag+cf\\
			de+bh&dg+bf
			\end{array}\right).
			$$ 
			Compare this with the iterant multiplication.
			$$([a,b] + [c,d]\eta)([e,f]+[g,h]\eta) = $$
			$$[a,b][e,f] + [c,d]\eta[g,h]\eta + [a,b][g,h]\eta + [c,d]\eta[e,f] =$$
			$$[ae,bf] + [c,d][h,g] +( [ag, bh] + [c,d][f,e])\eta = $$
			$$[ae,bf] +[ch,dg] + ( [ag, bh] + [cf,de])\eta = $$
			$$[ae+ch, dg+bf] + [ag + cf, de+bh]\eta.$$
Thus matrix multiplication is identical with iterant multiplication. The concept of the iterant can be used to
motivate matrix multiplication.
\bigbreak

\noindent The four matrices that can be framed in the two-dimensional wave 
form are all obtained from the two iterants
 $[a,d]$ and $[b,c]$ via the shift operation $\eta [x,y] = [y,x] \eta$ which we 
shall denote by  an overbar as shown below  $$\overline{[x,y]} = [y,x].$$
  
\noindent Letting  $A = [a,d]$  and $B=[b,c]$, we see that the four matrices seen in the 
grid are $$A + B \eta, B + A \eta, \overline{B} + \overline{A}\eta,
  \overline{A} + \overline{B}\eta.$$

 \noindent The operator  $\eta$  has the effect of rotating an iterant by ninety 
 degrees in the formal plane. Ordinary matrix multiplication can be written in a
 concise form using the following rules:

 $$\eta \eta = 1$$
 $$\eta Q = \overline{Q} \eta$$  where  Q is any two element iterant.
 Note the correspondence
 $$                      \left(\begin{array}{cc}
			a&b\\
			c&d
			\end{array}\right)
			=
			\left(\begin{array}{cc}
			a&0\\
			0&d
			\end{array}\right) 
			\left(\begin{array}{cc}
			1&0\\
			0&1
			\end{array}\right)
		        +
		        \left(\begin{array}{cc}
			b&0\\
			0&c
			\end{array}\right)
			\left(\begin{array}{cc}
			0&1\\
			1&0
			\end{array}\right) 
			= [a,d]1 + [b,c]\eta.$$ 
This means that $[a,d]$ corresponds to a diagonal matrix.
			 $$[a,d] = 	\left(\begin{array}{cc}
			a&0\\
			0&d
			\end{array}\right),$$ 
			$\eta$ corresponds to the anti-diagonal permutation matrix.
 $$\eta = 	\left(\begin{array}{cc}
			0&1\\
			1&0
			\end{array}\right),$$ 
			and $[b,c]\eta$ corresponds to the product of a diagonal matrix and the permutation matrix. $$[b,c]\eta = 	\left(\begin{array}{cc}
			b&0\\
			0&c
			\end{array}\right)
			\left(\begin{array}{cc}
			0&1\\
			1&0
			\end{array}\right) =
			 \left(\begin{array}{cc}
			0&b\\
			c&0
			\end{array}\right).$$  
			Note also that 		
$$\eta [c,b] = 
\left(\begin{array}{cc}
			0&1\\
			1&0
			\end{array}\right)
				 \left(\begin{array}{cc}
			c &0\\
			0&b
			\end{array}\right) =
			 \left(\begin{array}{cc}
			0&b\\
			c&0
			\end{array}\right).$$
			This is the matrix interpretation of the equation
			$$[b,c]\eta = \eta [c,b].$$
			\bigbreak
			
The fact that the iterant expression $ [a,d]1 + [b,c]\eta$ captures the whole of $2 \times 2$ matrix algebra corresponds to the fact that a two by two matrix is
combinatorially the union of the identity pattern (the diagonal)  and the interchange pattern (the antidiagonal) that correspond to the operators $1$ and $\eta.$
$$\left(\begin{array}{cc}
			*&@\\
			@ & *\\
			\end{array}\right)$$
			In the formal diagram for a matrix shown above, we indicate the diagonal by $*$ and the anti-diagonal by $@.$
\bigbreak

In the case of complex numbers we represent 
$$\left(\begin{array}{cc}
			a&-b\\
			b&a
			\end{array}\right) = [a,a] + [-b,b]\eta = a1 + b[-1,1]\eta = a + bi.$$ 
			In this way, we see that all of $2 \times 2$ matrix algebra is a hypercomplex number system based on the symmetric group $S_{2}.$
			In the next section we generalize this point of view to arbirary finite groups.
			\bigbreak 
 
 \noindent We have reconstructed the square root of minus one in the form of 
the matrix
   $$ i = \epsilon \eta = [-1,1]\eta 
   =\left(\begin{array}{cc}
			0&-1\\
			1&0
			\end{array}\right).$$
In this way, we arrive at this well-known representation of the complex numbers in terms of matrices.
Note that if we identify  the ordered pair $(a,b)$ with $a +ib,$ then this means taking the identification
$$(a,b) =  \left(\begin{array}{cc}
			a&-b\\
			b&a
			\end{array}\right).$$ Thus the geometric interpretation of multiplication by $i$ as a ninety degree rotation in the Cartesian plane, 
			 $$i(a,b) = (-b,a),$$ takes the place of the matrix equation
$$ i (a,b) = \left(\begin{array}{cc}
			0&-1\\
			1&0
			\end{array}\right)
			 \left(\begin{array}{cc}
			a&-b\\
			b&a
			\end{array}\right)
			=  \left(\begin{array}{cc}
			-b&-a\\
			a&-b
			\end{array}\right) =b + ia = (-b,a).$$
In iterant terms we have $$i[a,b] = \epsilon \eta [a,b] = [-1,1] [b,a] \eta  = [-b,a] \eta,$$ and this corresponds to the matrix equation
$$ i [a,b] = \left(\begin{array}{cc}
			0&-1\\
			1&0
			\end{array}\right)
			 \left(\begin{array}{cc}
			a& 0\\
			0 &b
			\end{array}\right)
			=  \left(\begin{array}{cc}
			0&-b\\
			a& 0
			\end{array}\right) =[-b,a] \eta.$$ All of this points out how the complex numbers, as we have previously examined them, live naturally in the context of the non-commutative algebras of iterants and  matrices. The factorization of $i$ into a product $\epsilon \eta$ of non-commuting iterant operators is closer both to the temporal nature of $i$ and to its algebraic roots.
\bigbreak
			
\noindent  More generally, we see that 
 $$(A + B\eta)(C+D\eta) = (AC+B\overline{D}) + (AD + 
B\overline{C})\eta$$

 \noindent writing the $2 \times 2$ matrix algebra as a system of 
 hypercomplex numbers.   Note that 
 $$(A+B\eta)(\overline{A}-B\eta) = A\overline{A} - B\overline{B}$$
 
 \noindent The formula on the right equals the determinant of the 
 matrix. Thus we define the {\em conjugate} of
 $Z = A+B\eta$ by the formula 
$$\overline{Z} = \overline{A+B\eta} = \overline{A} - B\eta,$$ and we have the formula
$$D(Z) = Z \overline{Z}$$ for the determinant $D(Z)$ where
$$Z= A + B\eta = 
\left(\begin{array}{cc}
			a&c\\
			d&b
			\end{array}\right)$$ where $A=[a,b]$ and $B=[c,d].$ Note
			that $$A\overline{A} =[ab, ba] = ab1 = ab,$$ so that
			$$D(Z) = ab -cd.$$ Note also that we assume that $a,b,c,d$ are in a commutative base ring. 
	\bigbreak	
	Note also that for $Z$ as above, 
	$$\overline{Z} = \overline{A} - B\eta =
	\left(\begin{array}{cc}
			b&-c\\
			-d&a
			\end{array}\right).$$ This is the classical adjoint of the matrix $Z.$
			\bigbreak
		
			We leave it to the reader to check that for matrix iterants $Z$ and $W,$
			$$Z\overline{Z} = \overline{Z}Z$$ and that 
			$$\overline{ZW} = \overline{W}\overline{Z}$$ and 
			$$\overline{Z + W} = \overline{Z} + \overline{W}.$$
			Note also that $$\overline{\eta} = - \eta, $$ whence
			$$\overline{B\eta} = - B\eta = -\eta \overline{B} = \overline{\eta} \overline{B}.$$
			We can prove that 
			$$D(ZW) = D(Z)D(W)$$ as follows
			$$D(ZW) = ZW \overline{ZW} = ZW \overline{W} \,\overline{Z} =
			 Z \overline{Z}W \overline{W} = D(Z)D(W).$$
			 Here the fact that $W \overline{W}$ is in the base ring which is commutative allows us to remove it from in between the appearance of $Z$ and $\overline{Z}.$ Thus we see that 
			 iterants as $2 \times 2$ matrices form a direct non-commutative generalization of 
			 the complex numbers.
 \bigbreak

 It is worth pointing out the first precursor to the quaternions ( the so-called {\it split quaternions}): This 
 precursor is the system  $$\{\pm{1}, \pm{\epsilon}, \pm{\eta}, 
\pm{i}\}.$$
 Here $\epsilon\epsilon = 1 = \eta\eta$ while $i=\epsilon \eta$ so 
that $ii = -1$.  
 The basic operations in this
 algebra are those of epsilon and eta.  Eta is the delay shift operator 
that reverses
 the components of the iterant. Epsilon negates one of the 
components, and leaves the
 order unchanged. The quaternions arise directly from these two 
operations once
 we construct an extra
 square root of minus one that commutes with them. Call this extra
 root of minus one $\sqrt{-1}$. Then the quaternions are generated 
by 
 $$I=\sqrt{-1}\epsilon, J= \epsilon \eta, K= \sqrt{-1}\eta$$
  with $$I^{2} = J^{2}=K^{2}=IJK=-1.$$
 The ``right" way to generate the quaternions is to start at the bottom 
iterant level
 with boolean values of $0$ and $1$ and the operation EXOR (exclusive or). Build 
iterants on this,
 and matrix algebra from these iterants.  This gives the square root 
of negation. Now
 take pairs of values from this new algebra and build $2 \times 2$ matrices 
again.  
 The coefficients include square roots of negation that commute with 
constructions at the
next level and so quaternions appear in the third level of this 
hierarchy. We will return to the quaternions after discussing other examples that involve matrices of all sizes.
\bigbreak

\section {Iterants of Arbirtarily High Period}
As a next example, consider a waveform of period three.
$$\cdots abcabcabcabcabcabc \cdots$$
Here we see three natural iterant views (depending upon whether one starts at $a$, $b$ or $c$).
$$[a,b,c],\,\,\, [b,c,a], \,\,\, [c,a,b].$$ The appropriate shift operator is given by the formula
$$[x,y,z]S = S[z,x,y].$$ Thus, with $T = S^{2},$
$$[x,y,z]T = T[y,z,x]$$ and $S^{3} = 1.$
With this we obtain a closed algebra of iterants whose general element is of the form $$[a,b,c] + [d,e,f]S + [g,h,k]S^{2}$$ where $a,b,c,d,e,f,g,h,k$ are real or complex numbers.
Call this algebra $\mathbb{V}ect_{3}(\mathbb{R})$ when the scalars are in a commutative ring with unit  $\mathbb{F}.$ Let $M_{3}(\mathbb{F})$ denote the $3 \times 3$ matrix algebra over $\mathbb{F}.$ We have the 
\smallbreak 

\noindent {\bf Lemma.} The iterant algebra  $\mathbb{V}ect_{3}(\mathbb{F})$ is isomorphic to the full
$3 \times 3$ matrix algebra $M_{3}((\mathbb{F}).$
\smallbreak

\noindent {\bf Proof.} Map $1$ to the matrix
$$\left(\begin{array}{ccc}
			1&0&0\\
			0&1&0\\
			0&0&1
			\end{array}\right).$$
Map $S$ to the matrix 
$$\left(\begin{array}{ccc}
			0&1&0\\
			0&0&1\\
			1&0&0
			\end{array}\right),$$
                          and map $S^2$ to the matrix
                           $$\left(\begin{array}{ccc}
			0&0&1\\
			1&0&0\\
			0&1&0
			\end{array}\right),$$
			Map $[x,y,z]$ to the diagonal matrix
			$$\left(\begin{array}{ccc}
			x&0&0\\
			0&y&0\\
			0&0&z
			\end{array}\right).$$
			Then it follows that $$[a,b,c] + [d,e,f]S + [g,h,k]S^{2}$$ maps to the matrix
			$$\left(\begin{array}{ccc}
			a&d&g\\
			h&b&e\\
			f&k&c
			\end{array}\right),$$ preserving the algebra structure. Since any $3 \times 3$ matrix can be written uniquely in this form, 
			it follows that  $\mathbb{V}ect_{3}(\mathbb{F})$ is isomorphic to the full
                          $3 \times 3$ matrix algebra $M_{3}(\mathbb{F}).$ $//$

We can summarize the pattern behind this expression of $3 \times 3$ matrices  by the following symbolic matrix.
$$\left(\begin{array}{ccc}
			1&S&T\\
			T&1&S\\
			S&T&1
			\end{array}\right)$$
			Here the letter $T$ occupies the positions in the matrix that correspond to the permutation matrix that represents it, and the letter $T = S^2$ occupies
			the positions corresponding to its permutation matrix.  The $1$'s occupy the diagonal for the corresponding identity matrix. The iterant representation
			corresponds to writing the $3 \times 3$ matrix as a disjoint sum of these permutation matrices such that the matrices themselves are closed under multiplication.
			In this case the matrices form a permutation representation of the cyclic group of order $3$, $C_{3} = \{1, S, S^{2} \}.$
			\bigbreak
			
\noindent {\bf Remark.} Note that a permutation matrix is a matrix of zeroes and ones such that some 
permutation of the rows of the matrix transforms it to the identity matrix. Given an $n \times n$ permutation matrix $P,$ we associate to it a permuation 
$$\sigma(P):\{1,2, \cdots, n\} \longrightarrow \{1,2, \cdots, n\}$$ via the following formula
$$i \sigma(P) = j$$ where $j$ denotes the column in $P$ where the $i$-th row has a $1$.
Note that an element of the  domain of a permutation is indicated to the left of the symbol for the permutation.
It is then easy to check that for permutation matrices $P$ and $Q$, 
$$\sigma(P)\sigma(Q) = \sigma(PQ)$$ given that we compose the permutations from left to right according to this convention.
\bigbreak

			It should be clear to the reader that this construction generalizes directly for iterants of any period and hence for a set of operators forming a cyclic group of any order.
			In fact we shall generalize further to any finite 
			group $G.$ We now define $\mathbb{V}ect_{n}(G,\mathbb{F})$ for any finite group $G.$
			\bigbreak
			
			\noindent {\bf Definition.} Let $G$ be a finite group, written multiplicatively. Let $\mathbb{F}$ denote a given commutative ring with unit. Assume that $G$ acts as a group of permutations on the set $\{1,2, 3,\cdots, n \}$ so that given an element $g \in G$ we have (by abuse of notation) $$g: \{1,2, 3,\cdots, n \} \longrightarrow \{1,2, 3,\cdots, n \}.$$
			We shall write $$ig$$ for the image of $i \in \{1,2, 3,\cdots, n \}$ under the permutation represented by $g.$ Note that this denotes functionality from the left and so we ask that $(ig)h = i(gh)$ for all elements $g, h \in G$ and $i1 = i$ for all $i$, in order to have a representation of $G$ as permutations. We shall call an $n$-tuple of elements of 
			$\mathbb{F}$ a {\it vector } and denote it by $a = (a_{1},a_{2},\cdots, a_{n}).$ We then define an action of $G$ on vectors over $\mathbb{F}$ by the formula
			$$a^{g} = (a_{1g}, a_{2g}, \cdots, a_{ng}),$$ and note that $(a^{g})^{h} = a^{gh}$ for all $g,h \in G.$ We now define an algebra  $\mathbb{V}ect_{n}(G,\mathbb{F})$, the 
			{\it iterant algebra for $G$,} to be the set of finite sums of formal products of vectors and group elements in the form $ag$ with multiplication rule
			$$(ag)(bh) = ab^{g}(gh),$$ and the understanding that $(a + b)g = ag + bg$ and for all vectors $a,b$ and group elements $g.$ It is understood that vectors are added 
			coordinatewise and multiplied coordinatewise. Thus $(a + b)_{i} = a_{i} + b_{i}$  and $(ab)_{i} = a_{i}b_{i}.$
			\bigbreak
			
			\noindent {\bf Theorem.} Let G be a finite group of order $n.$ Let $\rho: G \longrightarrow S_{n}$ denote the right regular representation of $G$ as permutations of
			$n$ things where we list the elements of $G$ as $G = \{ g_{1}, \cdots , g_{n}\}$ and let $G$ act on its own underlying set via the definition
			 $ g_{i} \rho(g) = g_{i}g.$  Here we describe $\rho(g)$ acting on the set of elements $g_{k}$ of $G.$ If we wish 
			 to regard $\rho(g)$ as a mapping of the set $\{1,2,\cdots n\}$ then we replace $g_{k}$ by $k$ and 
			$ i\rho(g) = k$ where $g_{i}g = g_{k}.$
			\smallbreak
			 Then $\mathbb{V}ect_{n}(G,\mathbb{F})$ is isomorphic to the matrix algebra $M_{n}((\mathbb{F}).$ In particular, we have that
			 $\mathbb{V}ect_{n!}(S_{n},\mathbb{F})$ is isomorphic with the matrices of size $n! \times n!$, $M_{n!}((\mathbb{F}).$
			 \smallbreak
			 
			 \noindent {\bf Proof.} See \cite{Kauffman5}.$//$\\
			 
\noindent {\bf Examples.} 
\begin{enumerate}
\item We have already implicitly given examples of this process of translation.
Consider the cyclic group of order three. $$C_{3} = \{1,S,S^2 \}$$ with $S^3 = 1.$ The multiplication table is 
$$\left(\begin{array}{ccc}
			1&S&S^2\\
			S&S^2&1\\
			S^2&1&S
			\end{array}\right).$$
			Interchanging the second and third rows, we obtain
			$$\left(\begin{array}{ccc}
			1&S&S^2\\
			S^2&1&S\\
			S&S^2&1
			\end{array}\right),$$ and this is the  $G$-{\it Table} that we used for 
			$\mathbb{V}ect_{3}(C_{3},\mathbb{F})$ prior to proving the Main Theorem.
			\smallbreak
			
			The same pattern works for abitrary cyclic groups. for example, consider the cyclic group of order $6.$ $C_{6} = \{1,S,S^2,S^3,S^4,S^5\}$ with $S^6 = 1.$ The multiplication table is
			$$\left(\begin{array}{cccccc}
			1&S&S^2&S^3&S^4&S^5\\
			S&S^2&S^3&S^4&S^5&1\\
			S^2&S^3&S^4&S^5&1&S\\
			S^3&S^4&S^5&1&S&S^2\\
			S^4&S^5&1&S&S^2&S^3\\
                            S^5&1&S&S^2&S^3&S^4\\
			\end{array}\right).$$
Rearranging to form the  $G$-{\it Table}, we have
$$\left(\begin{array}{cccccc}
			1&S&S^2&S^3&S^4&S^5\\
			 S^5&1&S&S^2&S^3&S^4\\
			 S^4&S^5&1&S&S^2&S^3\\
			 S^3&S^4&S^5&1&S&S^2\\
	                    S^2&S^3&S^4&S^5&1&S\\
			S&S^2&S^3&S^4&S^5&1\\
			\end{array}\right).$$
The permutation matrices corresponding to the positions of $S^{k}$ in the  $G$-{\it Table} give the matrix representation that gives the isomorphsm of $\mathbb{V}ect_{6}(C_{6},\mathbb{F})$ with the full algebra of 
six by six matrices.

\item In this example we consider the group $G = C_{2} \times C_{2},$ often called the ``Klein $4$-Group." We take $G = \{1,A,B,C\}$ where $A^2 = B^2 = C^2 = 1, AB = BA = C.$ Thus $G$ has the multiplication table, which is also its  $G$-{\it Table} for $\mathbb{V}ect_{4}(G,\mathbb{F}).$
$$\left(\begin{array}{cccc}
			1& A&B&C\\
			 A&1&C&B\\
			  B&C&1& A\\
			 C&B&A&1\\
			\end{array}\right).$$
			Thus we have the following permutation matrices that I shall call $E, A, B, C:$
			$$E = \left(\begin{array}{cccc}
			1& 0&0&0\\
			 0&1&0&0\\
			  0&0&1& 0\\
			 0&0&0&1\\
			\end{array}\right),
			A = \left(\begin{array}{cccc}
			0&1&0&0\\
			 1&0&0&0\\
			 0&0&0&1\\
			 0&0&1&0\\
			\end{array}\right),$$
			$$B = \left(\begin{array}{cccc}
			0&0&1&0\\
			 0&0&0&1\\
			 1&0&0&0\\
			 0&1&0&0\\
			\end{array}\right),
			C = \left(\begin{array}{cccc}
			0&0&0&1\\
			 0&0&1&0\\
			 0&1&0&0\\
			 1&0&0&0\\
			\end{array}\right).$$ The reader will have no difficulty verifying that 
			 $A^2 = B^2 = C^2 = 1, AB = BA = C.$
			 Recall that $[x,y,z,w]$ is iterant notation for the diagonal matrix
			 $$[x,y,z,w] = \left(\begin{array}{cccc}
			x&0&0&0\\
			 0&y&0&0\\
			 0&0&z&0\\
			 0&0&0&w\\
			\end{array}\right).$$ Let 
			$$\alpha = [1,-1,-1,1], \beta = [1,1,-1,-1], \gamma = [1,-1,1,-1].$$
			And let $$I = \alpha A, J = \beta B, K = \gamma C.$$
			Then the reader will have no trouble verifying that 
			$$I^2 = J^2 = K^2 = IJK = -1, IJ = K, JI = -K.$$
			Thus we have constructed the quaternions as iterants in relation to the Klein Four Group.
			in Figure~\ref{fourgroup} we illustrate these quaternion generators with string diagrams for the permutations. The reader can check that the permuations correspond to the permutation matrices constructed for the Klein Four Group. For example, the permutation for $I$ is $(12)(34)$ in cycle notation, the permutation for $J$ is $(13)(24)$ and the permutation for $K$ is $(14)(23).$ In the Figure
			we attach signs to each string of the permutation. These ``signed permutations'' act exactly as the products of vectors and permutations that we use for the iterants. One can see that the quaternions arise naturally from the Klein Four Group by attaching signs to the generating permutations as we have done in this Figure.

\begin{figure}
     \begin{center}
     \begin{tabular}{c}
     \includegraphics[width=6cm]{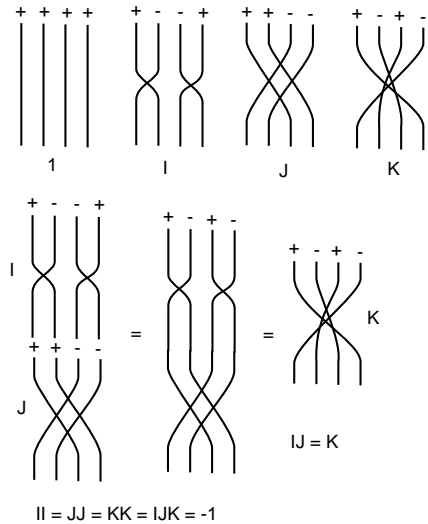}
     \end{tabular}
     \caption{\bf Quaternions From Klein Four Group}
     \label{fourgroup}
\end{center}
\end{figure}
\bigbreak

\item One can use the quaternions as a linear basis for $4 \times 4$ matrices just as our theorem would use the permutation matrices $1, A,B,C.$ If we restrict to real scalars $a,b,c,d$ such that $a^2 + b^2 + c^2 + c^2 = 1,$ then the set of matrices of the form $a1 + bI + cJ + dK$ is isomorphic to the group 
$SU(2).$
To see this, note that $SU(2)$ is the set of matrices with complex entries $z$ and $w$ with 
determinant $1$ so that $z \bar{z} + w \bar{w} = 1.$
$$M = \left(\begin{array}{cc}
			z& w\\
			 -\bar{w}&\bar{z}\\
			\end{array}\right).$$
Letting $z = a + bi$ and w = $c + di,$ we have
$$M =                 \left(\begin{array}{cc}
			a + bi& c + di\\
			 -c+di &a - bi\\
			\end{array}\right) $$
			$$= a
			\left(\begin{array}{cc}
			1& 0\\
			 0&1\\
			\end{array}\right) +b
			\left(\begin{array}{cc}
			i& 0\\
			 o&-i\\
			\end{array}\right) + c
			\left(\begin{array}{cc}
			0& 1\\
			 -1&0\\
			\end{array}\right)+d
			\left(\begin{array}{cc}
			0& i\\
			 i&0\\
			\end{array}\right).
			$$
If we regard $i = \sqrt{-1}$ as a commuting scalar, then we can write the generating matrices in terms
of size two iterants and obtain $$I=\sqrt{-1}\epsilon, J= \epsilon \eta, K= \sqrt{-1}\eta$$ as described in the previous section. IF we regard these matrices with complex entries as shorthand for $4 \times  4$ matrices with $i$ interpreted as a $2 \times 2$ matrix as we have done above, then these $4 \times 4$ matrices representing the quaternions are exactly the ones we have constructed in relation to the Klein Four Group.

Since complex numbers commute with one another, we could consider iterants whose values are in the complex numbers. This is just like considering matrices whose entries are complex numbers.
For this purpose we shall allow given  a version of $i$ that commutes with the iterant shift operator 
$\eta.$ Let this commuting $i$ be denoted
by $\iota.$ Then we are assuming that 

$$\iota^2 = -1$$ 
$$\eta \iota= \iota \eta$$
$$\eta^2 = +1.$$

We then consider iterant views of the form $[a + b\iota, c+ d\iota]$ and
$[a + b\iota , c + d\iota ]\eta = \eta[c + d\iota , a + b\iota ].$ In particular, we have
$\epsilon = [1,-1],$ and $i = \epsilon \eta$  is quite distinct from $\iota.$ Note, as before, that
$\epsilon \eta  = -\eta  \epsilon$ and that $\epsilon^2 = 1.$ Now let

$$I = \iota \epsilon$$ 
$$J = \epsilon \eta$$  
 $$K = \iota \eta.$$

We have used the commuting version of the square root of minus one in these definitions, and indeed we find the quaternions once more. 

$$I^2 =  \iota \epsilon \iota \epsilon =  \iota \iota \epsilon \epsilon = (-1)(+1) = -1,$$
$$J^2 =  \epsilon \eta \epsilon \eta =  \epsilon  (-\epsilon) \eta  \eta  = -1,$$
$$K^2 =   \iota \eta  \iota \eta=   \iota  \iota \eta \eta  = -1,$$
$$IJK = \iota \epsilon  \epsilon \eta \iota \eta  = \iota  1 \iota \eta \eta  = \iota \iota  = -1.$$

Thus
$$I^2 = J^2 = K^2 = IJK = -1.$$

This construction shows how the structure of the quaternions comes directly from the non-commutative structure of period two iterants. In other, words, quaternions can be represented by $2 \times  2$ matrices. This is the way it has been presented in standard language. The group $SU(2)$ of
$2 \times 2$  unitary matrices of determinant one is isomorphic to the quaternions of length one.

\item $$H = [a,b] + [c + d\iota , c-d\iota ]\eta =
\left(\begin{array}{cc}
			a&c + d\iota\\
			  c-d\iota &b\\
			\end{array}\right).$$

represents a Hermitian $2 \times 2$ 
matrix and hence an observable for quantum processes mediated by
$SU(2).$ Hermitian matrices have real eigenvalues.
\bigbreak

If in the above Hermitian matrix form we take $a=T+X, b=T-X, c = Y, d=Z,$ then we obtain an iterant and/or matrix representation for a point in Minkowski spacetime. 
 $$H = [T+X,T-X] + [Y + Z\iota , Y-Z\iota ]\eta $$
 $$= \left(\begin{array}{cc}
			T+X&Y + Z\iota\\
			  Y-Z\iota &T-X\\
			\end{array}\right).$$
Note that we have the formula $$Det(H) = T^2 - X^2 - Y^2 - Z^2.$$ It is not hard to see that the eigenvalues of $H$ are $T \pm \sqrt{X^2 + Y^2 + Z^2}.$ Thus, viewed as an observable, $H$ can observe the time and the invariant spatial distance from the origin of the event $(T,X,Y,Z).$ At least at this very elementary juncture, quantum mechanics and special relativity are reconciled.

\item Hamilton's Quaternions are generated by iterants, as discussed above, and we can express them
purely algebraicially by writing the corresponding permutations as shown below.

$$I = [+1,-1,-1,+1]s$$
$$J= [+1,+1,-1,-1]l$$
$$K= [+1,-1,+1,-1]t$$
where 

$$s =(12)(34)$$
$$l= (13)(24)$$
$$t =(14)(23).$$

Here we represent the permutations as products of transpositions $(ij).$ The transposition $(ij)$ interchanges $i$ and $j,$  leaving all other elements of $\{1,2,...,n \}$ fixed.

One can verify that 

$$I^2 = J^2 = K^2 = IJK = -1.$$
For example,

$$I^2 = [+1,-1,-1,+1]s [+1,-1,-1,+1]s$$
$$= [+1,-1,-1,+1][-1,+1,+1,-1]s s $$
$$= [-1,-1,-1,-1] $$
$$= -1.$$
and

$$IJ = [+1,-1,-1,+1]s [+1,+1,-1,-1]l$$
$$= [+1,-1,-1,+1][+1,+1,-1,-1] s l$$
$$= [+1,-1,+1,-1] (12)(34)(13)(24)$$
$$= [+1,-1,+1,-1] (14)(23)$$
$$= [+1,-1,+1,-1] t.$$

Nevertheless, we must note that making an iterant interpretation of
an entity like $I = [+1,-1,-1,+1]s$  is a conceptual departure from our original period two iterant (or cyclic period $n$) notion. Now we are considering iterants such as $[+1,-1,-1,+1]$ where the permutation group acts to produce other orderings of a given sequence. The iterant itself is not necessarily an oscillation. It can represent an implicate form that can be seen in any of its possible orders. These orders are subject to permutations that produce the possible views of the iterant. Algebraic structures such as the quaternions appear in the explication of such implicate forms.
\bigbreak

 The richness of the quaternions arises from the closed algebra that arises with its infinity of elements that satisfy the equation $U^2 = -1:$  $$U = aI + bJ + cK$$ where $a^2 + b^2 + c^2  = 1.$\\

\item In all these examples, we have the opportunity to interpret the iterants as short hand for matrix algebra based on permutation matrices, or as indicators of discrete processes. The discrete processes become more complex in proportion to  the complexity of the groups used in the construction.
We began with processes of order two, then considered cyclic groups of arbitrary order, then the symmetric group $S_{3}$ in relation to $6 \times 6$ matrices, and the Klein Four Group in relation to the quaternions. In the case of the quaternions, we know that this structure is intimately related to rotations of three and four dimensional space and many other geometric themes. It is worth reflecting on the possible significance of the underlying discrete dynamics for this geometry, topology and related physics.

\end{enumerate}

\section{The Framed Braid Group}
\begin{figure}
     \begin{center}
     \begin{tabular}{c}
     \includegraphics[width=6cm]{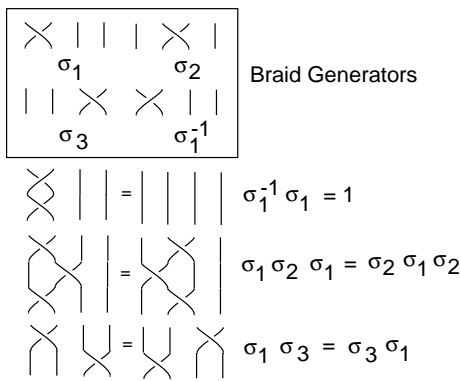}
     \end{tabular}
     \caption{\bf Braid Generators}
     \label{braidgen}
\end{center}
\end{figure}

The reader should recall that the symmetric group $S_{n}$ has presentation $$S_{n} = (T_{1},\cdots T_{n-1} | T_{i}^{2} = 1, T_{i}T_{i+1}T_{i} = T_{i+1}T_{i}T_{i+1}, T_{i}T_{j} = T_{j}T_{i}; |i-j|>1).$$
The Artin Braid Group $B_{n}$ is a relative of the symmetric group that is obtained by removing the condition that each generator has square equal to the idenity.
$$B_{n} = (\sigma_{1},\cdots \sigma_{n-1} | \sigma_{i}\sigma_{i+1}\sigma_{i} = \sigma_{i+1}\sigma_{i}\sigma_{i+1}, \sigma_{i}\sigma_{j} = \sigma_{j}\sigma_{i}; |i-j|>1).$$
In Figure~\ref{braidgen} we illustrate the the generators $\sigma_{1}, \sigma_{2}, \sigma_{3}$ of the $4$-strand braid group and we show the topological nature of the relation
$\sigma_{1}\sigma_{2}\sigma_{1} = \sigma_{2}\sigma_{1}\sigma_{2}$  and the commuting relation $\sigma_{1}\sigma_{3} = \sigma_{3}\sigma_{1}.$ Topological braids are represented as collections of 
always descending strings, starting from a row of points and ending at another row of points. The strings are embedded in three dimensional space and can wind around one another. The elementary braid 
generators $\sigma_{i}$ correspond to the $i$-th strand interchanging with the $i+1$-th strand. Two braids are multiplied by attaching the bottom endpoiints of one braid to the top endpoints of the other braid to form a
new braid.\\

There is a fundamental homomorphism $$\pi: B_{n} \longrightarrow S_{n}$$ defined on generators by $$\pi(\sigma_{i}) = T_{i}$$ in the language of the presentations above. In term of the diagrams in Figure~\ref{braidgen},
a braid diagram is a permutation diagram if one forgets about its weaving structure of over and under strands at a crossing.\\

\begin{figure}
     \begin{center}
     \begin{tabular}{c}
     \includegraphics[width=6cm]{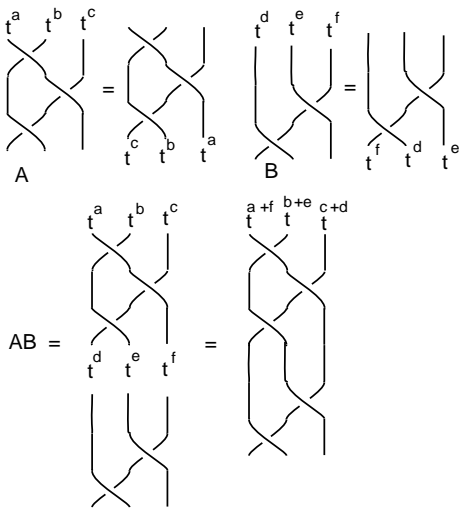}
     \end{tabular}
     \caption{\bf Framed Braids}
     \label{framedbraids}
\end{center}
\end{figure}

We now turn to a generalization of the braid group, the {\it framed braid group}. In this generalization, we associate elements of the form $t^{a}$  to the top of each braid strand. For these purposes it is useful to take $t$ as 
an algebraic variable and $a$ as an integer. To interpret this framing geometrically replace each braid strand by a ribbon and interpret $t^{a}$ as a $2\pi a$ twist in the ribbon. In Figure~\ref{framedbraids} we illustrate
how to multiply two framed braids. In our formalism the braids $A$ and $B$ in this figure are given by the formulas
$$A = [t^a, t^b,t^c] \sigma_{1}\sigma_{2}\sigma_{3},$$
$$B=[t^d,t^e,t^f] \sigma_{2}\sigma_{3}$$ in the framed braid group on three strands, denoted $FB_{3}.$ As the Figure~\ref{framedbraids} illustrates, we have the basic formula
$$ v \sigma = \sigma v^{\pi(\sigma)}$$ where $v$ is a vector of the form $v = [t^a, t^b,t^c]$ (for $n=3$) and $v^{\pi (\sigma)}$ denotes the action of the permutation associated with the braid $\sigma$ on the vector $v.$
In the figure the permutation is accomplished by sliding the algebra along the strings of the braid.\\

We can form an algebra $Alg[FB_{n}]$ by taking formal sums of framed braids of the form $\sum c_{k} v_{k} G_{k}$ where $c_{k}$ is a scalar, $v_{k}$ is a framing vector and $G_{k}$ is an element of the Artin Braid group $B_{n}.$ Since braids act on framing vectors by permutations, this algebra is a generalization of the iterant algebras we have defined so far. The algebra of framed braids uses an action of the braid group based on
its representation to the symmetric group. Furthemore, the representation $\pi: B_{n} \longrightarrow S_{n}$ induces a map of algebras 
$$\hat{\pi}: Alg[FB_{n}] \longrightarrow Alg[FS_{n}]$$ where we recognize $Alg[FS_{n}]$ as exactly an iterant algebra based in $S_{n}.$\\

\bigbreak

\bigbreak

\begin{figure}
     \begin{center}
     \begin{tabular}{c}
     \includegraphics[width=6cm]{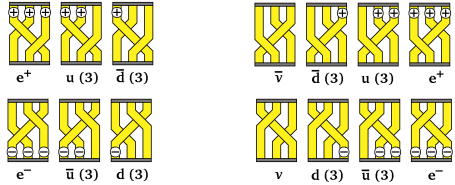}
     \end{tabular}
     \caption{\bf Sundance Bilson Thompson Framed Braid Fermions}
     \label{fermions}
\end{center}
\end{figure}

\begin{figure}
     \begin{center}
     \begin{tabular}{c}
     \includegraphics[width=6cm]{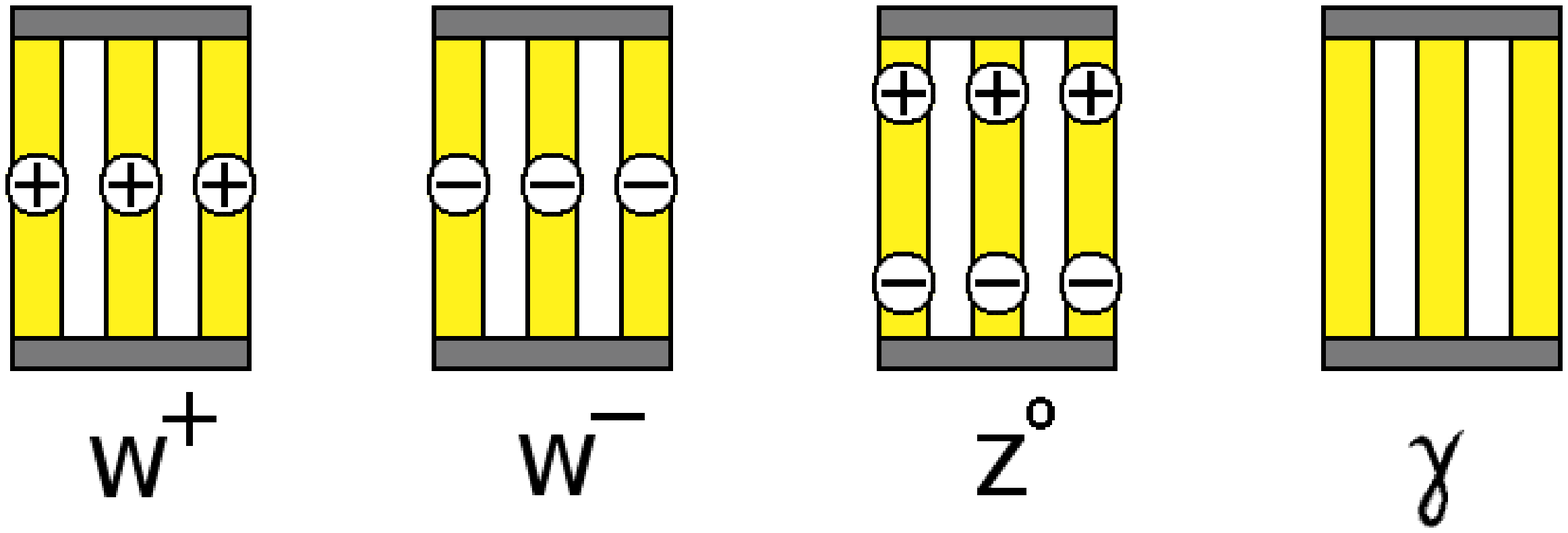}
     \end{tabular}
     \caption{\bf Bosons}
     \label{bosons}
\end{center}
\end{figure}

\begin{figure}
     \begin{center}
     \begin{tabular}{c}
     \includegraphics[width=6cm]{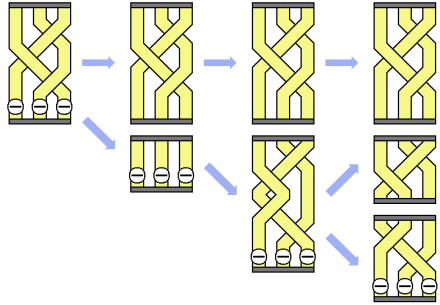}
     \end{tabular}
     \caption{\bf Representation of  $\mu \rightarrow \nu_\mu + W_{-} \rightarrow \nu_\mu  + \bar{\nu_e} +  e^{-}.$}
     \label{decay}
\end{center}
\end{figure}

In \cite{Sundance} Sundance Bilson-Thompson represents Fermions as framed braids. See Figure~\ref{fermions} for his diagrammatic representations.
In this theory each fermion is associated with a framed braid. Thus from the figure we see that the positron and the electron are given by the framed braids
$$e^{+} = [t,t,t]\sigma_{1} \sigma_{2}^{-1},$$ and
$$e^{-} = \sigma_{2} \sigma_{1}^{-1}[t^{-1},t^{-1},t^{-1}],$$
Here we use 
$[t^a, t^b,t^c]$ for the framing numbers $(a,b,c).$ Products of framed braids correspond to particle interactions. 
Note that $e^{+}e^{-}  = [1,1,1] = \gamma$ so that the electron and the positron are inverses in this algebra. In Figure~\ref{bosons} are illustrated the representations of bosons, including
$\gamma$, a photon and the identity element in this algebra. Other relations in the algebra correspond to particle interactions.
For example In Figure~\ref{decay} is illustrated the muon decay $$\mu \rightarrow \nu_\mu + W_{-} \rightarrow \nu_\mu  + \bar{\nu_e} +  e^{-}.$$ The reader can see the definitions of the
different parts of this decay sequence from the three figures we have just mentioned. Note the strictly speaking the muon decay is a multiplicative identity in the braid algebra:
$$\mu =  \nu_\mu W_{-} = \nu_\mu  \bar{\nu_e} e^{-}.$$ {\it Particle interactions in this model are mediated by factorizations in the non-commutative algebra of the framed braids.}\\

By using the representation
$\hat{\pi}: Alg[FB_{3}] \longrightarrow Alg[FS_{3}]$ we can image the structure of Bilson-Thompson's framed braids in the the iterant algebra corresponding to the symmetric group.
However, we propose to change this map so that we have a non-trivial representation of the Artin braid group. This can be accomplished by defining 
$$\rho: Alg[FB_{3}] \longrightarrow Alg[FS_{3}]$$ where
$$\rho(\sigma_{k}) = [t,t] T_{k}$$ and $$\rho(\sigma_{k}^{-1}) = [t^{-1},t^{-1}] T_{k}$$ for $k = 1,2.$ The reader will find that we have now mapped the braid group to the iterant algebra
$Alg[FS_{3}]$ and extended the mapping to the framed braid group algebra. Thus {\it the Sundance Bilson-Thompson representation of elementary particles as framed braids is mapped 
inside the iterant algebra for the symmetric group on three letters.} In Section 10 we carry this further and place the representation inside the Lie Algebra $su(3).$\\

\bigbreak

\section{Iterants and Parafermions}
Parafermions are the simplest generalization of Majorana fermions. As quasiparticles they exist in Read-Rezayi state in quantum Hall fluids\cite{Read}. Recently it has been shown that they exist as edge modes in parafermion chain models \cite{Fendley}in a similar way the Majorana fermions exist as edge modes in p-wave Kitaev chain\cite{Kitaev}. However parafermion algebra emerged in statistical models\cite{Fateev}. Interestingly parafermion algebra is a special case of the algebra Heisenberg-Weyl algebra\cite{Ortiz}.Parafermions are in focus in quantum computing research because they can be used to get the Fibonacci anyons which are universal for topological quantum computing.\\
In this section we take another route to iterant algebra using what we call as $Z_{n}$ clocks. We will write down the second order iterants in this way and  show how iterants are related to parefermions. At third order we will find that iterant algebra is same as parafermion algebra. We will also offer a new perspective of second order iterants.Central to our approach is the clock with roots of unity as time values. So right from the beginning we have cyclity and modular aithmetic built into the iterant algebra.

We begin with  a clock which has only two time values:two values of square root of unity. We call this clock a $Z_{2}$ clock. $Z_{2}$ here refers  to the set $\{1,-1\}$ which are values not only of square root of unity but can also refer to parity of Majorana fermion or to reflection of a spinor. In this section we will always refer to two values of square root of unity.
We introduce two operators or matrices. 
\begin{equation}
e=	\left(\begin{array}{cc}     
			1&0\\
			0&-1
			\end{array}\right)
\end{equation}
\begin{equation}
\eta=	\left(\begin{array}{cc}     
			0&1\\
			1&0
			\end{array}\right)
\end{equation}
We call $e$ as clock matrix and $\eta$ as shift matrix. Clock matrix has values of time as its eigenvalues while as shift matric changes the time states into one another and hence shifts the time on the $Z_{2}$ clock. 
$e$ and $\eta$ obey Clifford algebra. We can immediately see that $e$ and $\eta$ are actually the same as second order iterants which are obtained by starting from recursive processes present in imaginary unit,as discussed in section 5 of this paper. We identify that generators of iterant algebra are same(isomorphic) to the operators of our algebra. This new perspectives of iterant algebra proves useful for making higher gerenalization of the iterant algebra.


Lets now start with $Z_{3}$ clock which will have cube roots of unity as its time values:{1,$\omega$,$\omega^{2}$}. Then we can easily write down the clock and shift matrices.
\begin{equation}
e=	\left(\begin{array}{ccc}     
			1&0&0\\
			0&\omega&0\\
			0&0&\omega^{2}
			\end{array}\right)
\end{equation}
\begin{equation}
\eta=	\left(\begin{array}{ccc}     
			0&1&0\\
			0&0&1\\
			1&0&0
			\end{array}\right)
\end{equation}
But now $e$ and $\eta$ dont obey Clifford algebra. Rather they obey parafermion algebra.
\begin{equation}
e^{3}=\eta^{3}=1 \quad e\eta=\omega \eta e
\end{equation}





\section{ Iterants and the Standard Model}
In this section we shall give an iterant interpretation for the Lie algebra of the special unitary group $SU(3).$ The Lie algebra in question is denoted as $su(3)$ and is often described by a matrix basis.
The Lie algebra $su(3)$ is generated by the following eight Gell Man Matrices \cite{ChengLi}.
$$\lambda_{1} = \left(\begin{array}{ccc}
			0&1&0\\
			 1&0&0\\
			  0&0&0\\
			\end{array}\right),
\lambda_{2} = \left(\begin{array}{ccc}
			0&-i&0\\
			 i&0&0\\
			  0&0&0\\
			\end{array}\right),
\lambda_{3} = \left(\begin{array}{ccc}
			1&0&0\\
			 0&-1&0\\
			  0&0&0\\
			\end{array}\right),
			$$
$$\lambda_{4} = \left(\begin{array}{ccc}
			0&0&1\\
			 0&0&0\\
			  1&0&0\\
			\end{array}\right),
\lambda_{5} = \left(\begin{array}{ccc}
			0&0&i\\
			 0&0&0\\
			  -i&0&0\\
			\end{array}\right),
\lambda_{6} = \left(\begin{array}{ccc}
			0&0&0\\
			 0&0&1\\
			  0&1&0\\
			\end{array}\right),
			$$
$$\lambda_{7} = \left(\begin{array}{ccc}
			0&0&0\\
			 0&0&-i\\
			 0&i&0\\
			\end{array}\right),
\lambda_{8} = \frac{1}{\sqrt{3}}\left(\begin{array}{ccc}
			1&0&0\\
			 0&1&0\\
			 0&0&-2\\
			\end{array}\right)
			$$
			
The group $SU(3)$ consists in the matrices $U(\epsilon_{1},\cdots, \epsilon_{8}) = e^{i \sum_{a} \epsilon_{a} \lambda_{a}}$ where 
$\epsilon_{1},\cdots, \epsilon_{8}$ are real numbers and $a$ ranges from $1$ to $8.$ The Gell Man matrices satisfy the following relations.
$$tr(\lambda_{a} \lambda_{b}) = 2 \delta_{ab},$$
$$[\lambda_{a}/2 , \lambda_{b}/2] = i f_{abc} \lambda_{c}/2.$$
Here we use the summation convention -- summing over repeated indices, and $tr$ denotes standard matrix trace, $[A,B] = AB - BA$ is the matrix commutator and $\delta_{ab}$ is the Kronecker delta,
equal to $1$ when $a=b$ and equal to $0$ otherwise. The structure coefficients $f_{abc}$ take the following non-zero values.
$$f_{123} = 1,f_{147} = 1/2,f_{156} = -1/2,f_{246} = 1/2,f_{257} = 1/2,$$
$$f_{345} = 1/2,f_{367} = -1/2,f_{458} = \sqrt{3/2},f_{678} = \sqrt{3/2}$$

We now give an iterant representation for these matrices that is based on the pattern
$$ \left(\begin{array}{ccc}
			1&A&B\\
			 B&1&A\\
			 A&B&1\\
			\end{array}\right)$$ as described in the previous section. That is, we use the cyclic group of order three to represent all $3 \times 3$ matrices at iterants based on the permutation matrices
$$A = \left(\begin{array}{ccc}
			0&1&0\\
			 0&0&1\\
			 1&0&0\\
			\end{array}\right),
B= \left(\begin{array}{ccc}
			0&0&1\\
			 1&0&0\\
			 0&1&0\\
			\end{array}\right).$$ 
Recalling that $[a,b,c]$ as an iterant, denotes a diagonal matrix 
$$[a,b,c] = \left(\begin{array}{ccc}
			a&0&0\\
			 0&b&0\\
			 0&0&c\\
			\end{array}\right),$$  the reader will have no difficulty verifying the following formulas for the Gell Mann Matrices in the iterant format:
$$\lambda_{1} = [1,0,0]A + [0,1,0]B$$
$$\lambda_{2} = [-i,0,0]A + [0,i,0]B$$
$$\lambda_{3} = [1,-1,0]$$
$$\lambda_{4} = [1,0,0]B + [0,0,1]A$$
$$\lambda_{5} = [i,0,0]B + [0,0,-i]A$$
$$\lambda_{6} = [0,1,0]A + [0,0,1]B$$
$$\lambda_{7} = [0,-i,0]A + [0,0,i]B$$
$$\lambda_{8} = \frac{1}{\sqrt{3}}[1,1,-2].$$

Letting $F_{a} = \lambda_{a}/2,$ we can now rewrite the Lie algebra into simple iterants of the form $[a,b,c]G$ where $G$ is a cyclic group element. 
Compare with \cite{Gasiorowicz}.
Let
$$T_{\pm} = F_{1} \pm iF_{2},$$
$$U_{\pm} = F_{6} \pm i F_{7},$$
$$V_{\pm} = F_{4} \pm i F_{5},$$
$$T_{3} = F_{3},$$
$$Y = \frac{2}{\sqrt{3}}F_{8}.$$
Then we have the specific iterant formulas
$$T_{+} = [1,0,0]A, $$ $$T_{-} = [0,1,0]B,$$
$$U_{+} = [0,1,0]A, $$ $$U_{-} = [0,0,1]B,$$
$$V_{+} = [0,0,1]A, $$ $$V_{-} = [1,0,0]B,$$
$$T_{3} = [1/2,-1/2,0],$$
$$Y =  \frac{1}{\sqrt{3}}[1,1,-2].$$

We have that  $A[x,y,z] = [y,z,x]A$ and $B = A^{2} = A^{-1}$ so that $B[x,y,z] = [z,y,x]B.$ Thus we have reduced the basic $su(3)$ Lie algebra to a very elementary patterning of order three 
cyclic operations. In a subsequent paper, we will use this point to view to examine the irreducible representations of this algebra and to illuminate the Standard Model's Eightfold Way.\\

\section{ Iterants, Braiding and the Sundance-Bilson Thompson Model for Fermions}
In the last section we based our iterant representations on the following patterns and matrices.
The pattern,
$$ \left(\begin{array}{ccc}
			1&A&B\\
			 B&1&A\\
			 A&B&1\\
			\end{array}\right),$$ using the cyclic group of order three to represent all $3 \times 3$ matrices at iterants based on the permutation matrices
$$A = \left(\begin{array}{ccc}
			0&1&0\\
			 0&0&1\\
			 1&0&0\\
			\end{array}\right),
B= \left(\begin{array}{ccc}
			0&0&1\\
			 1&0&0\\
			 0&1&0\\
			\end{array}\right).$$ 
Recalling that $[a,b,c]$ as an iterant, denotes a diagonal matrix 
$$[a,b,c] = \left(\begin{array}{ccc}
			a&0&0\\
			 0&b&0\\
			 0&0&c\\
			\end{array}\right).$$ 
In fact there are six $3 \times 3$ permuation matrices: $\{ I, A, B, P,Q, R \}$ where
$$P = \left(\begin{array}{ccc}
			0&1&0\\
			 1&0&0\\
			 0&0&1\\
			\end{array}\right),
Q= \left(\begin{array}{ccc}
			1&0&1\\
			 0&0&1\\
			 0&1&0\\
			\end{array}\right), 
R= \left(\begin{array}{ccc}
			0&0&1\\
			 0&1&0\\
			 1&0&0\\
			\end{array}\right).$$ 
We then have $A = QP, B = PQ , R = PQP = QPQ.$  The two transpositions $P$ and $Q$ generate the entire group of permuatations $S_{3}.$ 
It is usual to think of the order-three transformations $A$ and $B$ as expressed in terms of these transpositons, but we can also use the iterant structure of the $3 \times 3$ matrices to
express $P,$ $Q$ and $R$ in terms of $A$ and $B.$ The result is as follows:
$$P = [0,0,1] + [1,0,0]A + [0,1,0]B,$$
$$Q = [1,0,0] + [0,1,0]A  + [0,0,1]B,$$
$$R= [0,1,0] + [0,0,1]A + [1,0,0]B.$$
Recall from the previous section that we have the iterant generators for the $su(3)$ Lie algebra:
$$T_{+} = [1,0,0]A, $$ $$T_{-} = [0,1,0]B,$$
$$U_{+} = [0,1,0]A, $$ $$U_{-} = [0,0,1]B,$$
$$V_{+} = [0,0,1]A, $$ $$V_{-} = [1,0,0]B.$$
Thus we can express these transpositions $P$ and $Q$ in the iterant form of the Lie algebra as
$$P = [0,0,1] + T_{+} + T_{-},$$
$$Q = [1,0,0] + U_{+}  + U_{-},$$
$$R = [0,1,0] + V_{+} + V_{-}.$$
The basic permutations receive elegant expressions in the iterant Lie algebra.\\

Now that we have basic permutations in the Lie algebra we can take the map from section 7.1
 $$\rho: Alg[FB_{3}] \longrightarrow Alg[FS_{3}]$$ with
$$\rho(\sigma_{k}) = [t,t] T_{k}$$ and $$\rho(\sigma_{k}^{-1}) = [t^{-1},t^{-1}] T_{k}$$ for $k = 1,2$
and send $T_{1}$ to $P$ and $T_{2}$ to $Q$. Then we have 
 $$\rho(\sigma_{1}) = [t,t]P$$ and $$\rho(\sigma_{1}^{-1}) = [t^{-1},t^{-1}]P$$ 
 and
 $$\rho(\sigma_{2}) = [t,t]Q$$ and $$\rho(\sigma_{1}^{-1}) = [t^{-1},t^{-1}]Q.$$ 
By choosing $t \ne 1$ on the unit circle in the complex plane, we obtain representations of the Sundance Bilson-Thompson constructions of Fermions via framed braids {\it inside} the 
$su(3)$ Lie algebra. This brings the Bilson-Thompson formalism in direct contact with the Standard Model via our iterant representations. We shall return to these relationships in a sequel to the present paper.\\

\section{Summary}
In this paper we have studied how quantum idempotents and the iterant algebras can generate essentially all fundamental algebras in physics. We have given representations of $su(2)$ algebra of spin,$su(3)$ algebra of quarks, Grassmann algebra of fermions,Clifford algebra of Majorana fermions and Heisenberg-Weyl algebra of parafermions. We have given a new representation of fermion algebra in terms of quantum idempotents and a new iterant algeba representation of the framed braid algebra of the Sundance Bilson-Thompson model and the $su(3)$ Lie algebra.  These representations are not only important in that they can give insights into new physics but also because they offer interpretations of spin, fermions and anyons. One very important aspect of these representations is that fermion algebra is presented without any need for second quantization and similarly spin algebra is also presented without any need for relativistic formalism. One of the very important findings is that all the algebras such as Lie algebras, Grassmann algebra,Clifford algebra are present in the algebra of quantum idempotents and naturally emerge from the basic iterant algebras. Since quantum idempotents encode quantum dynamics this shows that these algebras are algebras of quantum dynamics. Another aspect is that since Clifford algebras are geometric algebras so it also brings geometry closer to quantum dynamics. The way in which the framed braid algebra of the Sundance Bilson-Thomposn model for elementary particles fits into the
iterant representation and then is embedded into the $su(3)$ Lie algebra is remarkable, and will be the subject of further investigation.\\
\section{Acknowledgements}
Rukhsan Ul Haq would like to thank Department of Science and Technology(DST),India for funding and JNCASR Bangalore for a very conducive environment for research.

\end{document}